\documentclass[11pt]{article}
\usepackage{epsf}

\begin{document}

\begin{center}
{\large\bf Exact Analysis of Level-Crossing Statistics for (d+1)-Dimensional
Fluctuating Surfaces}\\

A. Bahraminasab,$^{1,2}$ M. Sadegh Movahed,$^{2,3}$ S. D.
Nasiri,$^4$ A. A. Masoudi,$^5$ Muhammad Sahimi$^6$

\end{center}

\noindent {\it $^1$International Center for Theoretical Physics, Strada
Costiera 11, I-34100 Trieste, Italy\\
$^2$Deptartment of Physics, Sharif University of Technology, Tehran 11365-9161,
Iran\\
$^3$Institute for Studies in Theoretical Physics and Mathematics, Tehran
19395-5531, Iran\\
$^4$Shahid Beheshti University, Tehran 19395-4716, Iran\\
$^5$Department of Physics, Alzahra University, Tehran 19834, Iran\\
$^6$Mork Family Department of Chemical Engineering \& Materials Science,
University of Southern California, Los Angeles, California 90089-1211, USA}

\bigskip

\noindent{\bf Abstract:}
We carry out an exact analysis of the average frequency $\nu_{\alpha x_i}^+$
in the direction $x_i$ of positive-slope crossing of a given level $\alpha$
such that, $h({\bf x},t)-\bar{h}=\alpha$, of growing surfaces in spatial
dimension $d$. Here, $h({\bf x},t)$ is the surface height at time $t$, and
$\bar{h}$ is its mean value. We analyze the problem when the surface growth
dynamics is governed by the Kardar-Parisi-Zhang (KPZ) equation without surface
tension, in the time regime prior to appearance of cusp singularities (sharp
valleys), as well as in the random deposition (RD) model. The total number
$N^+$ of such level-crossings with positive slope in all the directions is then
shown to scale with time as $t^{d/2}$ for both the KPZ equation and the RD
model.

\bigskip

\noindent PACS number(s): 52.75.Rx, 68.35.Ct.

\newpage

\noindent{\bf 1. Introduction}

\bigskip
Due to their practical applications and fundamental interest, a
great amount of effort has been devoted to understanding the
mechanism(s) of growth of thin films, and the kinetic roughening
of their surface during the growth. Analytical as well as
numerical analyses of many models of surface growth suggest, quite
generally, that certain properties of the surface exhibit dynamic
scaling.$^{1-6}$ In order to derive quantitative information about
the films' surface morphology, one considers a sample of size $L$
and defines the mean height $\bar{h}$ of the growing film and its
surface width $w$ by,$^1$
\begin{equation}
\bar{h}(L,t)=\frac{1}{L}\int_{-L/2}^{L/2} d{\bf x}\;h({\bf x},t)\;,
\end{equation}
where $h({\bf x},t)$ is the surface height, and
\begin{equation}
w(L,t)=\left[\langle(h-\bar{h})^2\rangle\right]^{1/2}\;,
\end{equation}
with $\langle\cdot\rangle$ indicating an average over different realizations
of the surface. Starting from a flat surface as the initial condition, it was
proposed by Family and Vicsek$^7$ that rescaling the space and time variables
by, respectively, $b$ and $b^z$ rescales the surface width $w$ by $b^\chi$,
$w(bL,b^zt)=b^\chi w(L,t)$, implying that
\begin{equation}
w(L,t)=L^\chi f(t/L^z)\;,
\end{equation}
where $f(x)$ is a scaling function, and $z$ is the dynamic exponent. If for
large $t$ and fixed $L$ ($t/L^z\to \infty$) the width $w$ saturates, then one
must have, $f(x)\to$ constant as $x\to\infty$. However, for fixed and large $L$
and $1\ll t\ll L^z$, one expects the correlations in the height fluctuations to
grow within only a distance $t^{1/z}$ and, thus, $w$ must be independent of
$L$. This implies that for $x\ll 1$, $f(x)\sim x^\beta$ with, $\beta=\chi/z$.
Therefore, the dynamic scaling of Family and Vicsek$^7$ postulates that,
$w(L,t)\sim t^\beta$ for $1\ll t\ll L^z$, and $w\sim L^\chi$ for $t\gg L^z$.
The roughness exponent $\chi$ and the dynamic exponent $z$ characterize the
self-affine geometry of the surface and its dynamics, respectively. These
scaling relations have been tested for a variety of growing surfaces by
extensive numerical simulations and analytical calculations.

In this paper we introduce the concept of level crossing in the context of
surface growth processes. In the level-crossing analysis one is interested in
determining the average frequency, $\nu_{\alpha x_i}^+$, in the $x_i-$direction
of observing a given level $\alpha$ for the function, $h-\bar{h}=\alpha$, in
the growing film. We show that $\nu_{\alpha x_i}^+$ is written in terms of the
joint probability distribution function (PDF) of $h-\bar{h}$ and the gradient
of $h({\bf x},t)$. Therefore, $\nu_{\alpha x_i}^+$ carries the same information
about the film's surface which is contained in the joint PDF of the
fluctuations in the height $h({\bf x},t)$ and its gradient. Our goal in this
paper is to study and analyze the quantity $\nu_{\alpha x_i}^+$ at time $t$ for
a growing surface in a sample of size L.

In addition, we introduce an integrated quantity, $N_{x_i}^+$, defined by,
$N_{x_i}^+=\int_{-\infty}^\infty d\alpha\;\nu_{\alpha x_i}^+$, which measures
the total number of positive-slope crossings of the surface in the
$x_i-$direction, and is expected to become size-dependent in the stationary
state. We determine exactly the time- and height-dependence of
$\nu_{\alpha x_i}^+$ and $N_{x_i}^+$ for the Kardar-Parisi-Zhang (KPZ) equation
in the limit of strong coupling, over time scales prior to the emergence of
cusp singularities (sharp valleys). We also derive exact expressions for the
same quantities for the random deposition model.

The statistics of level-crossings that we study in the present paper are
closely related to the concepts of first passage and persistence of
fluctuating interfaces in both space and time, which have recently been
studied in many papers.$^{8,9}$ In particular, the concept of spatial
persistence was introduced by Majumdar and Bray$^{10}$ (MB), who showed that
the probability $P_0(l)$ that the height of a fluctuating $(d+1)-$dimensional
surface in its steady state stays above its initial value up to a distance
$l$, along any linear cut in the $d-$dimensional space, decays as, $P_0(l)\sim
l^{-\theta}$. Here, $\theta$ is a spatial persistence exponent which takes on
distinct values, depending on how the point from which the distance $l$ is
measured is specified. Their analysis was carried out for a fluctuating
interface at {\it steady state}. The exponents were shown to map onto the
corresponding temporal persistence exponents for a generalized
$(d+1)-$dimensional random walk.

There are a few important differences between our work and that of MB. First,
all the analysis and calculations that we present in the present paper
are valid at times $t<t^*$, where $t^*$ is the time scale over which sharp
valleys (cusp singularities) are developed in the growing surface. Therefore,
the time regime that we investigate in the present paper is completely
different from what was considered by MB, as we study the probability of
crossing a certain height with a positive slope at times $t<t^*$, for which we
study the {\it short-time} behavior of the KPZ equation. Secondly, we study the
KPZ model in the limit of zero surface tension - a nonlinear equation - and
derive our exact results for any spatial dimension $d$, whereas MB derived
their results for a linear model (the Gaussian model), except when they
considered the KPZ equation in (1+1)-dimensions. Thirdly, we study the problem
assuming a flat initial surface, which is distinct from "finite initial
starting point" of MB (see below).

The plan of this paper is as follows. In Section 2 we discuss the connection
between $\nu_{\alpha x_i}^+$ and the underlying PDF of growing surfaces. In
Section 3 we derive an integral representation for $\nu_{\alpha x_i}^+$ for
the KPZ equation in $(d+1)-$dimensions in the strong coupling limit, before the
emergence of the cusp singularities. Section 4 presents our exact results for
the random deposition model. The paper is summarized in Section 5, while four
Appendices provide the details of the derivation of our results.

\newpage
\noindent{\bf 2. Level-Crossing in Growing Surfaces}

\bigskip
Consider a sample of an ensemble of functions which make up the homogeneous
random process $h({\bf x},t)$, representing the height of a growing surface.
Let $n_\alpha^+$ denote the number of positive-slope crossings such that,
$h({\bf x},t)-\bar{h}=\alpha$. In a time $t$ and for a typical growing surface
of linear size $L$ (see Figure $1$), let the mean value of $n_\alpha^+$ for all
the samples be $N_\alpha^+(L)$,
\begin{equation}
N_\alpha^+(L)=E[n_\alpha^+(L)]\;,
\end{equation}
where $E$ denotes the expectation or mean value of the quantity. If we consider
$N_\alpha^+$ in a second segment of size $L$ immediately following the first,
then, since the process is homogeneous, we obtain the same result as given by
Eq. (4). Thus, for the two intervals together we obtain,
\begin{equation}
N_\alpha^+(2L)=2N_\alpha^+(L)\;,
\end{equation}
from which it follows that, for a homogeneous process, the average number of
crossings is proportional to the space interval $L$. Hence,
\begin{equation}
N_\alpha^+(L)\propto L\;,
\end{equation}
or
\begin{equation}\label{y2}
N_\alpha^+(L)=\nu^+_\alpha L\;.
\end{equation}

We now consider how $\nu_\alpha^+$ is deduced from the underlying probability
distribution for $h({\bf x},t)-\bar{h}$. Consider a small length $dl$ of a
typical sample function. Since we assume that the process $h({\bf x},t)
-\bar{h}$ is a smooth function of {\bf x} with no sudden increase or decrease,
and that $dl$ is small enough, then the sample can only cross the level
$h({\bf x},t)-\bar{h}=\alpha$ with a positive slope, if $h({\bf x},t)-\bar{h}<
\alpha$ at the beginning of the interval $dl$. Furthermore, there is a minimum
slope at position {\bf x} if the level $h({\bf x},t)-\bar{h}=\alpha$ is to be
crossed in the interval $dl$, depending on the value of $h({\bf x},t)-\bar{h}$
at {\bf x}. Therefore, there will be a positive crossing of
$h({\bf x},t)-\bar{h}=\alpha$ in the next space interval $dl$, if at position
{\bf x},
\begin{equation}
h({\bf x},t)-\bar{h}<\alpha\;,\;\;\; {\rm and}\;\;\; \frac{d(h-\bar{h})}{dl}>
\frac{\alpha-[h({\bf x},t)-\bar{h}]}{dl}\;.
\end{equation}
If the above conditions are satisfied, then there will be a high probability of
crossing the level in the interval $dl.^{11,12}$

In order to determine whether conditions (8) are satisfied at an arbitrary
location {\bf x}, we must determine how values of $y=h({\bf x},t)-\bar{h}$
and $y'=dy/dl$ are distributed by considering their joint probability density
$P(y,y')$. Suppose that the level $y=\alpha$ and the interval $dl$ are
specified. Then, we are interested only in those values of $y$ such that
$y<\alpha$ and of $y'=dy/dl>(\alpha-y)/dl$, which represent the region between
the lines $y=\alpha$ and $y'=(\alpha-y)/dl$ in the plane ($y,y'$). Hence, the
probability of a positive-slope crossing of $y=\alpha$ in the interval $dl$ is
given by,
\begin{equation}\label{y1}
\int_0^\infty dy'\int_{\alpha-y'dl}^\alpha dy\; P(y,y^\prime)\;.
\end{equation}
As $dl\to 0$, one has,
\begin{equation}
P(y,y')=P(y=\alpha,y')\;.
\end{equation}
Since for large values of $y$ and $y'$ the PDF approaches zero fast enough,
Eq. (9) may be written as
\begin{equation}
\int_0^\infty dy'\int_{\alpha-y'dl}^\alpha dy\;P(y=\alpha,y')\;,
\end{equation}
in which the integrand is no longer a function of $y$, so that the first
integral in Eq. (11) is simply,\\$\int_{\alpha-y'dl}^\alpha dy\;P(y=\alpha,y')=
P(y=\alpha,y')y'dl$. Then, the probability of a positive-slope crossing of
$y=\alpha$ in the interval $dl$ is equal to
\begin{equation}
dl\int_0^\infty P(\alpha,y')y'dy'\;.
\end{equation}

Since according to Eq. (7) the average number of positive-slope crossings
over a length scale $L$ is $\nu_\alpha^+L$, then, the average number of
crossings in the interval $dl$ is $\nu^+_\alpha dl$. It then follows that the
average number of positive-slope crossings of the level $y=\alpha$ in the
interval $dl$ is equal to the probability of positive-slope crossing of the
level $y=\alpha$ in $dl$, which is true only if $dl$ is small and the process
$y(x)$ is smooth enough that there cannot be more than one crossing of
$y=\alpha$ in the interval $dl$. In that case, we have $\nu_\alpha^+dl=
dl\int_0^\infty P(\alpha,y')y'dy'$ and, therefore,
\begin{equation}\label{level}
\nu_\alpha^+=\int_0^\infty P(\alpha,y')y'dy'\;.
\end{equation}

In the following sections we will derive exact expressions for $\nu_\alpha^+$
via the joint PDF of $h({\bf x},t)-\bar{h}$ and the height gradient, for both
the KPZ and random deposition models. To derive the joint PDF we use the master
equation method.$^{13-16}$ This approach enables us to determine $\nu_\alpha^+$
in terms of a generating function. For example, the generating function for a
$(2+1)-$dimensional surface is given by (see below),
\begin{equation}
Z(\lambda,\mu,{\bf x},t)=\langle\exp\{-i\lambda[h({\bf x},t)-\bar{h}]-i\mu
u({\bf x},t)\}\rangle\;,
\end{equation}
where, $u({\bf x},t)=-\mbox{\boldmath$\nabla$}h$.

\bigskip
\noindent{\bf 3. Analysis of Level-Crossing for the KPZ Equation}

\bigskip
In the KPZ model in $(d+1)-$dimensions, the surface height $h({\bf{x}},t)$
at position {\bf x} on top of the substrate, in the limit of the zero
surface tension, satisfies the following stochastic equations
\begin{eqnarray}
& & \frac{\partial h}{\partial t}=h_t=\frac{1}{2}\bar{\alpha}\sum_{i=1}^d\
u_i^2+f \label{kp00}\;,\\
& & \frac{\partial u_i}{\partial t}=u_{i,t}=\bar{\alpha}\sum_{j=1}^d u_l
p_{ji}+f_{x_i}\label{kp20}\;,
\end{eqnarray}
where $u_i=\partial h/\partial x_i=h_{x_i}$, and $p_{ij}=
\partial h_{x_i}/\partial x_j$. Here, $f$ is a zero-mean random force with a
Gaussian correlation in space and white noise in time:
\begin{equation}
\langle f({\bf x},t)f({\bf x'},t')\rangle=2D_0D({\bf x}-{\bf x'})
\delta(t-t')\;,
\end{equation}
where $D({\bf x}-{\bf x}')$, an even function of its argument, is
the spatial correlation function which takes on the following form,
\begin{equation}
D({\bf x}-{\bf x'})=\frac{1}{\pi^{d/2}\sigma_{x_1}\sigma_{x_2}\cdots
\sigma_{x_d}}\exp \left[-\sum_1^d\frac{(x_i-x'_i)^2}{\sigma_{x_i}^2}\right]\;,
\end{equation}
with $\sigma_{x_i}$ being the standard deviations in the $x_i-$direction.
The parameters $\bar{\alpha}$ and $D_0$ describe, respectively, lateral growth
and the noise strength. Typically, in order to account for short-range
correlations, the correlation function is taken to be, $D({\bf x}-{\bf x}')=
\delta({\bf x}-{\bf x}')$, but we regularize the delta-function correlation by
a Gaussian function. When the standard deviations $\sigma=\sigma_{x_i}$ are
much smaller than the system's size $L$, we would expect the model to exhibit
short-range correlations in the $f$ term. Therefore, we emphasize that our
analysis is for the case when, $\sigma\ll L$.

As analyzed in Appendices A, B, and C, we define the generating function,
$Z(\lambda,\mu_i,x_i,t)=\langle\Theta(\lambda,\mu_i,x_i,t)\rangle$, for the
fields $\tilde{h}=h({\bf x},t)-\bar{h}$ and $u_i=h_{x_i}$, where
\begin{eqnarray}
\Theta=\exp\left\{-i\lambda\left[h({\bf x},t)-\bar{h}(t)\right]-i\sum_{i=1}^d
\mu_i u_i\right\}\;.
\end{eqnarray}
$\lambda$ and $\mu_i$ are the sources of $\tilde{h}$ and $u_i$, respectively,
and, $i,j=1,\cdots,d$. Assuming statistical homogeneity, i.e., assuming
that, $\partial Z/\partial{\bf x}=0$, it follows from Eqs. (15) and (16) that
$Z$ satisfies the following equation (see Appendix C for details),
\begin{equation}\label{zzm}
Z_t=\frac{\partial Z}{\partial t}=i\lambda\gamma(t)Z-\frac{1}{2}i\lambda
\bar{\alpha}\sum_lZ_{\mu_l\mu_l}-\lambda^2 k({\bf 0})Z\sum_l\mu_l^2
k''({\bf 0})Z\;.
\end{equation}
where, $k({\bf x}-{\bf x'})=2D_0D({\bf x}-{\bf x'})$, $\gamma(t)=\bar{h}_t$,
$k({\bf 0})=2D_0/(\pi\sigma^d)$, and $k_{x_ix_i}({\bf 0})=-4D_0/(\pi
\sigma^{2+d})$, where we use, $\sigma=\sigma_{x_i}=\sigma_{x_j}$, for
simplicity.

In trying to develop a statistical theory of level-crossings in rough surfaces,
it becomes clear that the interdependence of the statistics for height
difference $h({\bf x},t)-\bar{h}$ and height gradient must be taken into
account. The very existence of a nonlinear term in the KPZ equation leads to
development of the cusp singularities (sharp valleys) in a {\it finite time}
and in the strong coupling limit, hence forcing one to distinguish between
different time regimes. It was shown recently that, starting from a flat
surface, the KPZ equation will develop sharp-valley singularities after a time
scale $t^*$ where,$^{13}$ $t^*\sim {D_0}^{-1/3}\bar{\alpha}^{-2/3}
\sigma^{(d+4)/3}$. This implies that for times $t<t^*$ the relaxation
contributions tend to vanish in the strong coupling limit. In this regime one
can derive closed-form solution for the generating function. Starting from a
flat surface, i.e., from $h({\bf x},0)=0$ and $u({\bf x},0)=0$, one has the
following solution (see Appendix C)
\begin{equation} \label{ff}
Z(\lambda,\mu_1,\cdots,\mu_d,t)= F_1(\lambda,\mu_1,t)\cdots
F_d(\lambda,\mu_d,t)\exp[-\lambda^2 k({\bf 0})t]\;,
\end{equation}
with
\begin{displaymath}
F_j(\lambda,\mu_j,t)=\left\{1-\tanh^2\left[\sqrt{2ik_{xx}({\bf 0})\bar{\alpha}
\lambda}t\right]\right\}^{-1/4}
\end{displaymath}
\begin{equation}\label{24}
\times\exp\left\{-\frac{1}{2}i\mu_j^2\sqrt{\frac{2i
k_{xx}({\bf 0})}{\bar{\alpha}\lambda}}\tanh\left[\sqrt{2i{{k}_{{xx}}}({\bf 0})
\bar{\alpha}\lambda}t\right]-\frac{1}{2}i\bar{\alpha}k_{xx}({\bf 0})\lambda t^2
\right\}\;.
\end{equation}

To derive a closed expression for the PDF $P(\tilde{h},{\bf
u}_i,t)$, one needs to determine such moments as $\langle h^n
u_i^m u_j^l p_{ij}\rangle$. As shown in Appendix B, such moments
are identically zero.$^{14}$ Using this result, it can then be
shown that $P(\tilde{h},{\bf u}_i,t)$ takes on the following
expression (see Appendix C),
\begin{equation}\label{pp1}
P(\tilde{h},u_{x_i},t)=\frac{1}{(2\pi)^{d+1}}\int d\lambda d\mu_1\cdots
d\mu_d\;\exp\left(i\lambda\tilde{h}+i\sum_{i=1}^d\mu_iu_i\right)\;
Z(\lambda,\mu_1,\cdots\mu_d,t)\;.
\end{equation}
According to the Eq. (13), the frequency of crossing a definite
height $h({\bf x},t)-\bar{h}=\alpha$ with a positive slope in, for
example, the $x_1-$direction, is given by,
\begin{equation}
\nu_{\alpha x_1}^+=\int_0^\infty du_{x_1}\;u_{x_1}\int_{-\infty}^\infty
du_{x_2}\cdots du_{x_d}\; P(\alpha,u_{x_1},\cdots,u_{x_1},t)\;.
\end{equation}
Using the Eqs. (21)-(23), $\nu_{\alpha x_1}^+$ is then given by,
\begin{eqnarray} \label{ll}
& & \nu_{\alpha x_1}^+=\frac{1}{4\pi^2}\int_{-\infty}^\infty
\int_{-\infty}^\infty-\frac{e^{i\lambda\alpha}}{\mu_1^2}\;
Z(\lambda,\mu_1,\mu_2\to 0,\cdots,\mu_d\to 0,t)d\lambda d\mu_1,\nonumber\\
&=&\frac{1}{2\pi^{3/2}}\int^\infty_{-\infty}d\lambda\exp\left[i\lambda\alpha-
\lambda^2k({\bf 0})t-i\bar{\alpha}k_{xx}({\bf 0})\lambda t^2\right]\;\zeta,
\end{eqnarray}
where
\begin{equation}\label{zeta}
\zeta=\frac{\displaystyle\sqrt{\frac{1}{2}i\sqrt{\frac{2ik_{xx}(
{\bf 0})} {\bar{\alpha}\lambda}}\;\tanh\left[t\sqrt{2ik_{xx} ({\bf
0})\bar{\alpha}\lambda}
\right]}}{\displaystyle\sqrt{1-\tanh^2\left[t\sqrt{2ik_{xx}({\bf 0})
\bar{\alpha}\lambda}\right]}}\;.
\end{equation}

To compute the integral in Eq. (25), we used the numerical integration software
developed by Piessens {\it et al.}$^{17}$ which uses a globally-adaptive
scheme based on Gauss-Kronrod quadrature rules. In Fig. 2 we plot
$\nu_{\alpha x_1}^+$ for times, $t/t^*=0.05\;,0.1$, and $0.2$, which are before
the emergence of the cusp singularities. To derive an expression for
$N_{x_1}^+$, the total number of level-crossings with positive slopes in the
$x_1-$directions, let us express it in terms of the generating function $Z$. It
can be shown straightforwardly that $N_{x_1}^+$ is written in terms of the
generating function $Z$ as
\begin{equation}
N_{x_1}^+=\frac{1}{2\pi}\int_{-\infty}^\infty-\frac{1}{\mu_1^2}
Z(\lambda\to 0,\mu_1,\mu_2\to 0,\cdots,\mu_d\to 0,t)\;d\mu_1
=\lim_{\lambda\to 0}\frac{\zeta}{\sqrt{\pi}}\;.
\end{equation}
Using the Eq. (26) one finds that, $N_{x_1}^+\sim t^{1/2}$. In Fig. 3 (obtained
by direct numerical integration of $N_{x_1}^+=\int^\infty_{-\infty}\nu_{\alpha
x_1}^+\;d\alpha$) we plot $N_{x_1}^+$ vs $t$, which also indicates that,
$N_{x_1}^+\sim t^{1/2}$, in agreement with the analytical prediction.

\bigskip
\noindent{\bf 4. Analysis of Level-Crossing for the Random Deposition Model}

\bigskip
In the random deposition model, the height of each column performs an
independent random walk. This model leads to unrealistically rough surfaces,
with its width growing with the exponent $\beta=\chi/z=1/2$ without ever
saturating. In the continuum limit the random deposition model is described by,
\begin{equation}
\frac{\partial}{\partial t} h({\bf x},t)=f({\bf x},t)\;,\;\;\;
\frac{\partial}{\partial t} u_i({\bf x},t)=f_{x_i}\;,\;\;
i=1,\cdots,d\;,
\end{equation}
where, $u_i({\bf x},t)=\partial h({\bf x},t)/\partial x_i$, and $f({\bf x},t)$
is a zero-mean random force described by Eqs. (17) and (18). Assuming
homogeneity, we define the generating function by
\begin{equation}
Z(\lambda,\mu_i,t)=\left\langle\exp\left[-i\lambda h({\bf x},t)-i\sum_{i=1}^d
\mu_iu_i\right]\right\rangle\;.
\end{equation}
As before, $\lambda$ and $\mu_i$ are the sources of $\tilde{h}$ and $u_i$,
respectively, and $i,j=1,\cdots,d$.

It is then straightforward to derive the following equation for the evolution
of $Z(\lambda,\mu_i,t)$:
\begin{equation}
\frac{\partial}{\partial t}Z(\lambda,\mu,t)=-\lambda^2D_0D(0)Z+\sum_{i=1}^d
\mu_i^2D_0D_{x_ix_i}(0)Z\;.
\end{equation}
The joint PDF of $h$ and $u_i$ is obtained by Fourier transform of the
generating function:
\begin{equation}
P(h,u,t)=\frac{1}{2\pi}\int d\lambda d\mu_i\;\exp\left(i\lambda h+i\sum_{i=1}^d
\mu_i u_i\right)\;Z(\lambda,\mu_i,t)\;.
\end{equation}
Carrying out the Fourier transformation, we obtain a Fokker-Planck (FP)
equation,
\begin{equation}
\frac{\partial}{\partial t}P=D_0D(0)\frac{\partial^2}{\partial h^2}P-
\sum_{i=1}^dD_0D_{x_ix_i}(0)\frac{\partial^2}{\partial u_i^2}P\;.
\end{equation}
The solution of the above FP equation is written as, $P(h,{\bf
u}_i,t)= p_1(h,t)p_2(u_1,t)\cdots p_{d+1}(u_d,t)$(for motivation
see$^{20}$). Using the initial conditions that,
$p_1(h,0)=\delta(h)$, and, $p_i(u_i,0)=\delta(u_i)$, and starting
from a flat surface, it can then be shown that,
\begin{equation}
P(h,{\bf u}_i,t)=\frac{1}{(2\pi tD_0)^{(d+1)/2}[-D''(0)]^{d/2}}
\exp\left[-\frac{h^2}{4D_0D(0)t}+\frac{\sum_{i=1}^du_i^2}{4D_0D_{xx}(0)t}
\right]\;,
\end{equation}
where, $D''=D_{x_ix_i}=D_{x_jx_j}$. Thus, the frequency of crossing a
definite height, $h({\bf x},t)=\alpha$, in, for example, the $x-$direction,
is given by,
\begin{eqnarray}
\nu_{\alpha x}^+&=&\int_0^\infty u_xP(\alpha,u_x,u_j)du\nonumber \\
&=&\frac{1}{2\pi}\sqrt{-\frac{D_{xx}(0)}{D(0)}}\exp\left[
-\frac{\alpha^2}{4D(0)t}\right]=\frac{1}{2\pi\sigma}
\exp\left[-\frac{\alpha^2}{4D_0D(0)t}\right]\;,\;\;\;j=2,\cdots,d\;.
\end{eqnarray}
Thus, the quantity $\nu_{\alpha x}^+$ in the random deposition model has a
Gaussian form with respect to $\alpha$. Moreover, it is easily seen that
$N_x^+\sim t^{1/2}$, in agreement with the result for the KPZ equation.

\bigskip
\noindent{\bf 5. Summary}
\bigskip

In this paper we derived exact results for the statistics of level-crossing
with positive slopes for growing surfaces that are governed by the KPZ equation
in $(d+1)-$dimensions with a Gaussian forcing term which is $\delta-$correlated
in time and contains short-range spatial correlations, in the limit of zero
surface tension. The integral representation of the frequency of crossing,
$\nu_{\alpha x_i}^+$, in the $x_i-$direction is given for the KPZ equation in
the strong coupling limit before the emergence of sharp valleys (cusp
singularities). We also determined the quantity $N_{x_i}^+=
\int_{-\infty}^{+\infty }d\alpha\;\nu_{\alpha x_i}^+$, which measures the total
number of positive-slope crossing of the growing surface in the
$x_i-$direction, and showed that, $N_{x_i}^+\sim t^{1/2}$. Using statistical
homogeneity, it is then clear that, $\nu_{\alpha x_i}^+=\nu_{\alpha x_j}^+$,
and that, $N^+=N_{x_1}^+\cdots N_{x_d}^+\sim t^{d/2}$. We also derived exact
expressions for similar properties in the random deposition model.

The ideas and techniques presented in this paper are quite general and may be
used to determine $\nu_{\alpha x_i}^+$ for processes that are governed by a
general Langevin equation with given drift and diffusion coefficients.

\bigskip
\noindent{\bf Acknowledgment}

\bigskip
We would like to thank M. Reza Rahimi Tabar for useful comments.

\newpage

\begin{center}
{\bf Appendix A}
\end{center}

In this and the following Appendices we derive the governing equation for the
generating function, Eq. (20). In particular, in the present Appendix we
investigate a more general definition of the generating function in
$(d+1)$-dimensions, and will further expand the study in Appendix B in order to
derive an identity which is crucial for deriving Eq. (20) in Appendix C.

We define a generating function by
\begin{equation}
Z(\lambda,\mu_i,\eta_{ij},x_i,t)=
\langle\Theta(\lambda,\mu_i,\eta_{ij},x_i,t)\rangle\;,
\end{equation}
for the fields, $\tilde{h}=h-\bar{h}$, $u_i=h_{x_i}$, and $p_{ij}=h_{x_ix_j}$.
$\lambda,\;\mu_i$, and $\eta_{ij}$ are the sources of $\tilde{h}$, $u_i$ and
$p_{ij}$, respectively, and $i,j=1,\cdots,d$. The explicit expression for
$\Theta$ is given by,
\begin{equation}
\Theta=\exp\left\{-i\lambda\left[h({\bf x},t)-\bar{h}(t)\right]
-i\sum_{i=1}^d\mu_iu_i-i\sum_{i\leq
j=1}^d{\eta_{ij}p_{ij}}\right\}\;.
\end{equation}
We also define $q_{ijk}$ by, $q_{ijk}=h_{x_ix_jx_k}$, which will be used later.
Considering the zero-surface tension limit of the KPZ equation. The time
evolution of the height $h({\bf x},t)$ and its derivatives are given by,
\begin{eqnarray} \label{kp0}
&& h_t=\frac{1}{2}\bar{\alpha}\sum_{i=1}^d u_i^2+f,\label{kp0}\\
&& u_{i,t}=\bar{\alpha}\sum_{l=1}^d u_l p_{li}+f_{x_i},\label{kp2}\\
&& p_{ij,t}=\bar{\alpha}\sum_{l=1}^d p_{li}p_{lj}+\bar{\alpha}\sum_{l=1}^d
u_l q_{lij}+f_{x_ix_j}\;.\label{kp3}
\end{eqnarray}

Using Novikov's theorem$^{18,19}$ (see Appendix D) we can write down
the following identities,
\begin{eqnarray}\label{n1}
&& \langle f\Theta\rangle=-i\lambda k({\bf
0})Z-i\sum_{l=1}^{d}\eta_{ll}
k''({\bf 0})Z\;,\\
&& \langle f_{x_i}\Theta\rangle=-i\mu_ik''({\bf 0})Z\;,\\
&& \langle f_{x_ix_i}\Theta\rangle=-i\lambda k''({\bf
0})Z-i\sum_{l=1}^d
\eta_{ll}k''''({\bf 0})Z\;,\\
&& \langle f_{x_ix_j}\Theta\rangle=-i\eta_{ij}k''''({\bf 0})Z\hspace{1cm}i\neq
j\;,
\end{eqnarray}
where, for example, in $d=3$ we have
\begin{eqnarray}
&& k({\bf x}-{\bf x}')=2D_0D({\bf x}-{\bf x}')\;,\nonumber\\
&& k({\bf 0})=k({\bf 0})=\frac{2D_0}{(\pi\sigma^2)^{3/2}}\;,\nonumber\\
&& k'({\bf 0})=k_x({\bf 0})=k_y({\bf 0})=k_z({\bf
0})=0\;,\nonumber\\
&& k''=k_{xx}({\bf 0})=k_{yy}({\bf 0})=k_{zz}({\bf
0})=\frac{-4D_0}{\sigma^2(\pi\sigma^2)^{3/2}}\;.\nonumber
\end{eqnarray}
If we write down the above equations for $(d+1)$ dimensions, the only change
would be having $d$ components for the arguments appearing in $k$, $k''$, and
$k''''$.

Differentiating the generating function $Z$ with respect to $t$, and using Eqs.
(\ref{kp0}) - (43) and the following identity,
\begin{equation}
iZ_{x_l}-i\lambda Z_{\mu_l}-i\sum_{i}\mu_iZ_{\eta_{il}}\equiv\sum_{i\leq j}
\eta_{ij}\langle q_{ijl}\Theta\rangle\;,
\end{equation}
we find that the time evolution of $Z$ is governed by
\begin{displaymath}
Z_t=i\lambda\gamma(t)Z-i\frac{\lambda\bar{\alpha}}{2}\sum_lZ_{\mu_l\mu_l}
-i\bar{\alpha}\sum_lZ_{\eta_{ll}}+i\bar{\alpha}\sum_{l,i\leq j}\eta_{ij}
Z_{\eta_{li}\eta_{li}}-\lambda^2 k({\bf 0})Z+\sum_l\mu_l^2k''({\bf 0})Z
\end{displaymath}
\begin{equation}\label{zz}
-2\lambda\sum_{l}\eta_{ll}k''({\bf 0})Z-\left(\sum_{l,k}\eta_{ll}\eta_{kk}+
\sum_{l<k}\eta_{lk}^2\right)k''''({\bf 0})Z\;,
\end{equation}
where, $\gamma(t)=\bar{h}_t$. If we Fourier transform $Z$ with respect to
$\lambda,\;\mu_i$, and $\eta_{ij}$, we obtain the governing equation for the
joint PDF of $\tilde{h},\;u_i$, and $p_{ij}$,
\begin{displaymath}
P(\widetilde{h},u_i,p_{ij},t)=\int\frac{d\lambda}{2\pi}\prod_i
\frac{d\mu_i}{2\pi}\prod_{i\leq j}\frac{d\eta_{ij}}{2\pi}
\end{displaymath}
\begin{equation}\label{pp}
\times\exp\left\{i\lambda\left[h({\bf
x},t)+\bar{h}(t)\right]+i\sum_l\mu_lu_l +i\sum_{l\leq
k}{\eta_{lk}p_{lk}}\right\}Z(\lambda,\mu_i,\eta_{ij},x_i,t)\;.
\end{equation}
Using Eqs. (\ref{zz}) and (\ref{pp}), the governing equation for the time
evolution of $P(\tilde{h},u_i,p_{ij},t)$ in $d$ spatial dimensions is then
given by,
\begin{displaymath}
P_t=\gamma(t)P_{\tilde{h}}+\frac{\bar{\alpha}}{2}\sum_l
u_l^2P_{\tilde h}-\bar{\alpha} (d+2)\sum_lp_{ll}P
-\bar{\alpha}\sum_{l,k\leq m}p_{lk}p_{lm}P_{p_{km}}+
k({\bf 0})P_{\tilde{h}\tilde{h}}-k''({\bf 0})\sum_{l}P_{u_lu_l}\;,
\end{displaymath}
\begin{equation}\label{pt}
+2k''({\bf 0})\sum_lP_{\tilde{h}p_{ll}}+k''''({\bf 0})\sum_{l\leq k}P_{p_{lk}
p_{lk}}-2k''''({\bf 0})\sum_{l<k }P_{p_{ll}p_{kk}}\;.
\end{equation}
Equation (\ref{pt}) enables us to obtain the governing equations for the time
evolution of the moments of the height and its derivatives. The resulting
equation has the following general form,
\begin{displaymath}
\frac{\partial}{\partial t}\langle\tilde{h}^{n_0}
AB\rangle=-n_0\gamma(t)\langle\tilde{h}^{n_0-1}
AB\rangle-\frac{\bar{\alpha}n_0}{2}\sum_l\langle\tilde{h}^{n_0-1}
ABu_l^2\rangle+\bar{\alpha}\sum_{l,k\leq m}n_{km}\langle\tilde{h}^{n_0}
AB\frac{p_{lk}p_{lm}}{p_{km}}\rangle
\end{displaymath}
\begin{displaymath}
-\bar{\alpha}\sum_l\langle\tilde{h}^{n_0}ABp_{ll}\rangle+k({\bf 0})n_0(n_0-1)
\langle\tilde{h}^{n_0-2}AB\rangle - k''({\bf 0})\sum_ln_l(n_l-1)\langle
\frac{\tilde{h}^{n_0}AB}{u_l^2}\rangle+2k''({\bf 0})\sum_ln_0n_{ll}\langle
\frac{\tilde{h}^{n_0}AB}{p_{ll}^2}\rangle
\end{displaymath}
\begin{equation}\label{am}
+k''''({\bf 0})\sum_{l\leq k}n_{lk}(n_{lk}-1)\langle\frac{\tilde{h}^{n_0}AB}
{p_{lk}^2}\rangle + 2k''''({\bf 0})\sum_{l<k}n_{ll}n_{kk}\langle
\frac{\tilde{h}^{n_0}AB}{p_{ll}p_{kk}}\rangle\;,
\end{equation}
where
\begin{eqnarray}
&& A=\prod_{i=1}^du_i^{n_i}\;,\nonumber\\
&& B=\prod_{i\leq j}p_{ij}^{n_{ij}}\;.\nonumber
\end{eqnarray}
By using various values $n_0$, $n_i$, and $n_{ij}$, coupled differential
equations that govern the evolution of the moments are constructed.
\newpage

\begin{center}
{\bf Appendix B}
\end{center}

In this Appendix we prove the identity, $\langle p_{ij}\exp{(-i\lambda\tilde{h}
-i\sum_l\mu_lu_l)}\rangle=0$. As shown in Appendix C, this identity is crucial
to deriving  Eq. (20). We also determine the height moments exactly and show
that, $\langle h_{x_ix_j}\Theta\rangle=0$ for $i\neq j$, by considering an
initially flat surface. However, it can be shown that this identity also holds
for the general $(d+1)-$dimensional surfaces. Setting $n_0=n_i=n_{ij}=0$, we
obtain from Eq. (\ref{am}),
\begin{equation}
-\bar{\alpha}\sum_l\langle p_{ll}\rangle=-\bar{\alpha}\langle
\mbox{\boldmath$\nabla$}\cdot{\bf u}\rangle=0\Rightarrow
\langle\mbox{\boldmath$\nabla$}\cdot{\bf u}\rangle=0\;.
\end{equation}
The main aim is to calculate the moments $\langle p_{ij}\exp(-i\lambda\tilde{h}
-i\sum_l\mu_lu_l)\rangle$. In order to do so, we must follow several steps.
First, we must determine such moments as, $\langle p_{ij}u_iu_j\rangle$ for
$i\neq j$. Using Eq. (\ref{am}), we have
\begin{equation}\label{can}
\frac{\partial}{\partial t}\langle u_iu_jp_{ij}\rangle=\bar{\alpha}\sum_l
\langle u_iu_jp_{li}p_{lj}\rangle -\bar{\alpha}\sum_l\langle u_iu_jp_{ll}p_{ij}
\rangle=\bar{\alpha}\sum_l\langle u_iu_j(p_{li}p_{lj}-p_{ll}p_{ij})\rangle\;.
\end{equation}

Inspecting the right-hand side of Eq. (\ref{can}), we see that the terms with
$l=i$ or $l=j$ cancel one another. Noting that, for now, we have restricted
our attention to $(3+1)-$dimensions, Eq. (\ref{can}) is written as
\begin{equation}\label{can11}
\frac{\partial}{\partial t}\langle u_iu_jp_{ij}\rangle=\langle u_iu_j
(p_{li}p_{lj}-p_{ll}p_{ij})\rangle \hspace{.5cm}i \neq j\neq l\;.
\end{equation}
Inserting the right-hand side of Eq. (\ref{can11}) in Eq. (\ref{am}), one finds
\begin{displaymath}
\frac{\partial}{\partial t}\langle u_iu_j(p_{li}p_{lj}-p_{ll}p_{ij})
\rangle=\bar{\alpha}\sum_k\langle u_iu_jp_{ki}p_{kl}p_{lj}\rangle
+\bar{\alpha}\sum_k\langle u_iu_jp_{kj}p_{kl}p_{il}\rangle
\end{displaymath}
\begin{equation}\label{a}
-\bar{\alpha}\sum_k\langle u_iu_jp_{kk}p_{li}p_{lj}\rangle
-\bar{\alpha}\sum_k\langle u_iu_jp_{kl}p_{kl}p_{ij}\rangle
-\bar{\alpha}\sum_k\langle u_iu_jp_{ki}p_{kj}p_{ll}\rangle
+\bar{\alpha}\sum_k\langle u_iu_jp_{kk}p_{ij}p_{ll}\rangle\;.
\end{equation}
It can then be seen that for $i\neq j\neq l$ the right-hand side of
Eq. (\ref{a}) is zero which, when utilized, yields the following equation for a
flat initial surface,
\begin{equation}\label{a22}
\langle u_iu_j(p_{li}p_{lj}-p_{ll}p_{ij})\rangle=0\;,
\end{equation}
and
\begin{equation} \label{a23}
\langle u_iu_jp_{ij}\rangle=0\;.
\end{equation}
It can also be shown by induction that all the moments $\langle\tilde
h^{n_0}p_{ij}^{n_{ij}}u_i^{n_i}u_j^{n_j}\rangle$ are identically zero.
Therefore, we conclude that,
$\langle p_{ij}\exp{(-i\lambda\tilde h-i\sum_l\mu_lu_l)}\rangle=0$.

\newpage

\begin{center}
{\bf Appendix C}
\end{center}

Using the identities that we derived in Appendix B, we derive the
joint PDF for the KPZ equation in the limit of zero surface
tension. For clarity, we restrict ourselves to the
$(3+1)-$dimensional surface, but generalization of the results to
any $(d+1)-$dimensional surface is quite straightforward. The
zero-surface tension limit of the KPZ equation in
$(3+1)-$dimensions has the following form (for more clarity, we
replace the $h({\bf x},t)$ with the new notation $h(x,y,z,t)$ in
$(3+1)-$dimensions),
\begin{equation}\label{k1}
h_t(x,y,z,t)-\frac{1}{2}\bar{\alpha}(h_x^2+h_y^2+h_z^2)=f\;.
\end{equation}
Letting,
\begin{equation}
h_x=u,\;\;\; h_y=v,\;\;\; h_z=w\;
\end{equation}
and differentiating the KPZ equation with respect to $x$, $y$, and $z$, we have
\begin{displaymath}
u_t=\bar{\alpha}(uu_x+vv_x+ww_x)+f_x\;,
\end{displaymath}
\begin{displaymath}
v_t=\bar{\alpha}(uu_y+vv_y+ww_y)+f_y\;,
\end{displaymath}
\begin{equation}\label{r1}
w_t=\bar{\alpha}(uu_z+vv_z+ww_z)+f_z\;,
\end{equation}
for the height $h$ and the corresponding velocity fields. The generating
function $Z(\lambda,\mu_1,\mu_2,\mu_3,x,y,z,t)$ is defined so as to generate
the height and velocity field moments. By defining $\Theta$ as
\begin{equation}\label{t1}
\Theta=\exp\left\{-i\lambda[h(x,y,z,t)-\bar{h}(t)]-i\mu_1u-i\mu_2v-i\mu_3w
\right\}\;,
\end{equation}
the generating function is written as,
$Z(\lambda,\mu_1,\mu_2,\mu_3,x,y,z,t)=\langle\Theta\rangle$. Using the KPZ
equation and its derivatives, the time evolution of
$Z(\lambda,\mu_1,\mu_2,\mu_3,x,y,z,t)$ is then written as,
\begin{displaymath}
Z_t=i\gamma(t)\lambda
Z-\frac{1}{2}i\lambda\bar{\alpha}\langle(u^2+v^2+w^2)
\Theta\rangle-i\bar{\alpha}\mu_1\langle(uu_x+vv_x+ww_x)\Theta\rangle
-i\bar{\alpha}\mu_2\langle(uu_y+vv_y+ww_y)\Theta\rangle
\end{displaymath}
\begin{equation}\label{evolut}
-i\bar{\alpha}\mu_3\langle(uu_z+vv_z+ww_z)\Theta\rangle-i\lambda\langle
f\Theta\rangle-i\mu_1\langle f_x\Theta\rangle-i\mu_2\langle f_y\Theta\rangle
-i\mu_3\langle f_z\Theta\rangle\;,
\end{equation}
where $\gamma(t)=h_t=\frac{1}{2}\alpha\langle u^2+v^2+w^2\rangle$. Utilizing
the statistical homogeneity  of the quantities, we obtain
\begin{eqnarray}\label{h1}
Z_x & = &\langle(-i\lambda u-i\mu_1u_x-i\mu_2v_x-i\mu_3w_x)\Theta\rangle\;,\\
Z_y & = &\langle(-i\lambda v-i\mu_1u_y-i\mu_2v_y-i\mu_3w_y)\Theta\rangle\;, \\
Z_z & = &\langle(-i\lambda w-i\mu_1u_z-i\mu_2v_z-i\mu_3w_z)\Theta\rangle\;.
\end{eqnarray}

Because we analyze the system only in a time regime in which the cusp
singularities (sharp valleys) have not yet formed, the order of the partial
derivatives can be exchanged,
\begin{displaymath}
\frac{\partial^2 h}{\partial x_i\partial x_j}=\frac{\partial^2 h}
{\partial x_j\partial x_i}\;,
\end{displaymath}
and therefore, $v_x=u_y$, $w_x=u_z$, and $w_y=v_z$. Keeping the definition of
$\Theta$, Eq.(\ref{t1}), in mind, we may write
\begin{equation}\label{h2}
i\frac{\partial}{\partial\mu_1}\langle(-i\mu_1u_x-i\mu_2v_x-i\mu_3w_x)
\Theta\rangle=\langle u_x\Theta\rangle-i\mu_1\langle uu_x\Theta\rangle-i\mu_2
\langle uv_x\Theta\rangle-i\mu_3\langle uw_x\Theta\rangle\;.
\end{equation}
Using Eqs. (\ref{h1}) and (\ref{h2}) we obtain,
\begin{equation}
\langle u_x\Theta\rangle-i\mu_1\langle uu_x\Theta\rangle-i\mu_2\langle
uv_x\Theta\rangle-i\mu_3\langle uw_x\Theta\rangle=-\lambda\frac{\partial}
{\partial\mu_1}\langle u\Theta\rangle=-i\lambda Z_{\mu_1\mu_1}
\end{equation}
In a similar manner we find for the $y-$ and $z-$ directions that,
\begin{equation}
-i\lambda Z_{\mu_2\mu_2}=\langle v_y\Theta\rangle-i\mu_1\langle
vu_y\Theta\rangle-i\mu_2\langle vv_y\Theta\rangle-i\mu_3\langle vw_x\Theta
\rangle\;,
\end{equation}
\begin{equation}
-i\lambda Z_{\mu_3\mu_3}=\langle w_z\Theta\rangle-i\mu_1\langle
wu_z\Theta\rangle-i\mu_2\langle wv_z\Theta\rangle-i\mu_3\langle ww_z
\Theta\rangle\;.
\end{equation}
Using Novikov's theorem$^{18,19}$ (see Appendix D) the expressions
for $\langle f\Theta\rangle$ and $\langle f_{x_i}\Theta\rangle$ are written
in terms of $Z$ (see Appendix A). Therefore, we obtain
\begin{displaymath}
Z_t=i\gamma(t)\lambda Z-\frac{1}{2}i\lambda\bar{\alpha}Z_{\mu_1\mu_1}
-\frac{1}{2}i\lambda\bar{\alpha}Z_{\mu_2\mu_2}
-\frac{1}{2}i\lambda\bar{\alpha}Z_{\mu_3\mu_3}
\end{displaymath}
\begin{equation}
-\bar{\alpha}\langle u_x\Theta\rangle-\bar{\alpha}\langle v_y\Theta\rangle
-\bar{\alpha}\langle w_z\Theta\rangle-\lambda^2k({\bf 0},0)Z+\mu_1^2
k_{xx}({\bf 0},0)Z+\mu_2^2k_{xx}({\bf 0},0)Z+\mu_3^2k_{xx}({\bf 0},0)Z\;.
\end{equation}
The $\langle u_{x_i}\Theta\rangle$ terms are written as
\begin{displaymath}
\langle u_x\Theta\rangle=\frac{i}{\mu_1}\langle\Theta\rangle_x+\frac{i}
{\mu_1}\langle(i\lambda u+i\mu_2v_x+i\mu_3w_x)\Theta\rangle
\end{displaymath}
\begin{equation}
= -i\frac{\lambda}{\mu_1}Z_{\mu_1}-\frac{\mu_2}{\mu_1}\langle
v_x\Theta\rangle-\frac{\mu_3}{\mu_1}\langle w_x\Theta\rangle\;,
\end{equation}
and, similarly, for $\langle v_y\Theta\rangle$ and $\langle w_z\Theta\rangle$
we have
\begin{equation}
\langle v_y\Theta\rangle=-i\frac{\lambda}{\mu_2}Z_{\mu_2}
-\frac{\mu_1}{\mu_2}\langle u_y\Theta\rangle
-\frac{\mu_3}{\mu_2}\langle w_y\Theta\rangle\;,
\end{equation}
\begin{equation}
\langle w_z\Theta\rangle=-i\frac{\lambda}{\mu_3}Z_{\mu_3}
-\frac{\mu_1}{\mu_3}\langle u_z\Theta\rangle
-\frac{\mu_2}{\mu_3}\langle v_z\Theta\rangle\;.
\end{equation}
The terms $\langle h_{x_i x_j}\Theta\rangle$ $(i\neq j)$, that appear in
the equation for the time evolution of $Z$ (for example, $\langle u_y\Theta
\rangle$), prevent us from writing the $Z-$equation in a closed form.
Fortunately, as was shown in Appendix B, such terms are zero for a flat
initial condition. Therefore, the generating function $Z$ satisfies the
following equation
\begin{displaymath}
Z_t=i\gamma(t)\lambda Z-\frac{1}{2}i\lambda\bar{\alpha}Z_{\mu_1\mu_1}-
\frac{1}{2}i\lambda\bar{\alpha}Z_{\mu_2\mu_2}-\frac{1}{2}i\lambda\bar{\alpha}
Z_{\mu_3\mu_3}+i\bar{\alpha}\frac{\lambda}{\mu_1}Z_{\mu_1}-i\bar{\alpha}
\frac{\lambda}{\mu_2}Z_{\mu_2}
\end{displaymath}
\begin{equation}\label{evolut21}
-i\bar{\alpha}\frac{\lambda}{\mu_3}Z_{\mu_3}-\lambda^2k({\bf 0},0)Z
+\mu_1^2k_{xx}({\bf 0},0)Z+\mu_2^2k_{xx}({\bf 0},0)Z+\mu_3^2
k_{xx}({\bf 0},0)Z\;.
\end{equation}

We now solve Eq. (\ref{evolut21}) assuming a flat initial condition,
$h(x,y,z,0)=u(x,y,z,0)=v(x,y,z,0)=w(x,y,z,0)=0$, which is equivalent to writing
\begin{equation}\label{shart}
P(\tilde{h},u,v,0)=\delta(\tilde{h})\delta(u)\delta(v)\delta(w)\;,
\end{equation}
which means that
\begin{equation}
Z({0,\bf 0},t)=1\;.
\end{equation}
An efficient way of solving Eq. (\ref{evolut21}) is to factorize $Z$ in the
following manner,$^{11,15,16}$
\begin{equation}\label{factor}
Z(\lambda,\mu_1,\mu_2,\mu_3,t)=F_1(\lambda,\mu_1,t)F_2(\lambda,\mu_2,t)
F_3(\lambda,\mu_3,t)\exp[-\lambda^2 k({\bf 0})t]\;.
\end{equation}
Then, by inserting Eq. (\ref{factor}) in Eq. (\ref{evolut21}) we obtain
\begin{displaymath} \label{evolut3}
{F_1}_tF_2F_3+F_1{F_2}_tF_3+F_1F_2{F_3}_t=
i\gamma(t)\lambda F_1F_2F_3-\frac{1}{2}i\lambda\bar{\alpha}
F_2{F_1}_{\mu_1\mu_1}F_2F_3-\frac{1}{2}i\lambda\bar{\alpha}
F_1{F_2}_{\mu_2\mu_2}F_3
\end{displaymath}
\begin{displaymath}
-\frac{1}{2}i\lambda\bar{\alpha}F_1F_2{F_3}_{\mu_3\mu_3}
+i\bar{\alpha}\frac{\lambda}{\mu_1}{F_1}_{\mu_1}F_2F_3
-i\bar{\alpha}\frac{\lambda}{\mu_2}F_1{F_2}_{\mu_2}F_3
-i\bar{\alpha}\frac{\lambda}{\mu_3}F_1F_2{F_3}_{\mu_3}
-\lambda^2 k({\bf 0})F_1F_2F_2+\mu_1^2 k''({\bf 0})F_1F_2F_3
\end{displaymath}
\begin{equation}\label{evolut3}
+\mu_2 k''({\bf 0})F_1F_2F_{2}+\mu_3k''({\bf 0})F_1F_2F_2\;.
\end{equation}
Therefore,
\begin{equation}\label{f}
F_t=-\frac{1}{2}i\lambda\bar{\alpha}F_{\mu\mu}+
i\bar{\alpha}\frac{\lambda}{\mu}F_\mu+[\mu^2 k''({\bf 0})
-i\bar{\alpha}\lambda k''({\bf 0})t]F\;,
\end{equation}
with the initial condition, $F(\lambda,\mu,0)=1$. This indicates that the
height gradients in the three dimensions evolve independently of one another
before the emergence of cusp singularities, and that they are only coupled with
the height field. By Fourier transforming Eq. (\ref{f}) with respect to $\mu$
a simpler partial differential equation of order one is obtained, which can be
solved by the method of characteristics.$^{11}$ The solution of $F_j$ is then
given by
\begin{equation}
F_j(\mu,\lambda,t)=\left\{1-\tanh^2\left[t\sqrt{2ik_{xx}({\bf
0})\bar{\alpha} \lambda}\right]\right\}^{-1/4}
\exp\left\{-\frac{1}{2}i\bar{\alpha}k''({\bf 0})\lambda t^2
-\frac{1}{2}i\mu^2\sqrt{\frac{2ik_{xx}({\bf
0})}{\bar{\alpha}\lambda}} \tanh\left[t\sqrt{2ik_{xx}({\bf
0})\bar{\alpha}\lambda}\right]\right\}.
\end{equation}
Therefore, we obtain the generating function in $(3+1)-$dimensions,
\begin{displaymath}
Z(\lambda,\mu_1,\mu_2,t)=
\left\{1-\tanh^2\left[t\sqrt{2ik_{xx}({\bf
0})\bar{\alpha}\lambda}\right]
\right\}^{-3/4}\exp\left\{-\frac{1}{2}i(\mu_1^2+\mu_2^2+\mu_3^2)
\sqrt{\frac{2ik_{xx}({\bf
0})}{\bar{\alpha}\lambda}}\tanh\left[t\sqrt{2ik_{xx} ({\bf
0})\bar{\alpha}\lambda}\right]\right\}
\end{displaymath}
\begin{equation}\label{h}
\times\exp\left[-\frac{3}{2}i\bar{\alpha}k''({\bf 0})\lambda t^2-k({\bf 0})
\lambda^2t\right]\;.
\end{equation}
By inverse Fourier transformation of the generating function $Z$, the PDF of
the height fluctuations, $P(\tilde{h},u,v,w,t)$, is then determined.

As mentioned earlier, generalizing the above results to any $(d+1)-$dimensional
surface is straightforward. For example, for $d=2$ all we need to do is setting
$\mu_3=0$ in Eq. (\ref{h}). For a general $d$ we have the following expression
for $Z$:
\begin{equation}\label{factor}
Z(\lambda,\mu_1,\cdots,\mu_d,t)=F_1(\lambda,\mu_1,t)\cdots
F_d(\lambda,\mu_d,t)\exp[-\lambda^2 k({\bf 0})t]\;,
\end{equation}
where $F_j$ is defined as before.

\newpage

\begin{center}
{\bf Appendix D}
\end{center}

In this Appendix we prove a simple version of the theorem by Novikov$^{18,19}$
that we use in this paper. As before, we define the generating,
$Z(\lambda,\mu_i,x_i,t)=\langle\Theta(\lambda,\mu_i,x_i,t)\rangle$, for the
fields $\tilde{h}=h({\bf x},t)-\bar{h}$ and $u_i=h_{x_i}=
\partial h/\partial x_i$ by
\begin{equation}
\Theta=\exp\left\{-i\lambda[h(x,y,z,t)-\bar{h}(t)]-i\sum_{i=1}^d\mu_iu_i
\right\}\;,
\end{equation}
where, as usual, $\lambda$ and $\mu_i$ are the sources of $\tilde{h}$ and
$u_i$, respectively, and $i,j=1,\cdots,d$. Starting with the equation,
\begin{equation}
\partial_t h({\bf x},t)=-{\cal L}[h({\bf x},t)]+f({\bf x},t)
\end{equation}
where $f({\bf x},t)$ is a term representing the noise with Gaussian
correlations in space and white noise in time:
\begin{equation}
\langle f({\bf x},t)f({\bf x}',t')\rangle=2D_0D({\bf x}-{\bf x}')
\delta(t-t')\;,
\end{equation}
we can write,
\begin{equation}
\langle F[f]f({\bf x},t)\rangle=\int d{\bf x}'dt'\;\langle f({\bf
x},t) f({\bf x}',t')\rangle\left\langle\frac{\partial F[f]}{\partial
f({\bf x}',t')}\right\rangle\;.
\end{equation}
If we take $F=\Theta$, we obtain
\begin{equation}
\frac{\partial F[f]}{\partial f({\bf x}',t')}=-i\lambda
\frac{\partial h({\bf x},t)}{\partial f({\bf x}',t')}F[f]
-i\sum_{i=1}^d\mu_i\frac{\partial u_i({\bf x},t)} {\partial f({\bf
x}',t')}F[f]\;.
\end{equation}
Using Eq. (81) we can write,
\begin{equation}
h({\bf x},t)=h({\bf x},t_0)-\int_{t_0}^t{\cal L}[h({\bf x},t)]dt'+
\int_{t_0}^tf({\bf x},t')dt'
\end{equation}
which, when differentiated with respect to $f$, gives,
\begin{equation}
\frac{\partial h({\bf x},t)}{\partial f({\bf x},t)}=-\int_{t_0}^t
\frac{\partial{\cal L}[h({\bf x},t)]}{f({\bf x},t)}dt' +\delta({\bf
x}-{\bf x}')\;.
\end{equation}
Using Eqs. (84) and (86) in the limit, $t-t'\to 0$, invoking
casuality, and considering the same expressions for,
$\sum_{i=1}^d\mu_i\partial u_i({\bf x},t)/\partial f({\bf
x}',t')F[f]$, we obtain
\begin{equation}
\langle F[f]f({\bf x},t)\rangle=-2i\lambda D_0D(0)\langle F[f]
\rangle-2i\sum_{i=1}^d\mu_iD_0D''({\bf x}-{\bf x}')\langle
F[f]\rangle\;.
\end{equation}
Using the last expressions and $k({\bf x})=2D_0D({\bf x})$, we obtain Eq.
(\ref{n1}).

\newpage

\noindent{\bf References}

\begin{enumerate}

\item A.-L. Barab\'asi and H. E. Stanley, {\it Fractal Concepts in Surface
Growth} (Cambridge University Press, New York, 1995).

\item T. Halpin-Healy and Y. C. Zhang, Phys. Rep. {\bf 245}, 218 (1995); J.
Krug, Adv. Phys. {\bf 46}, 139 (1997).

\item J. Krug and H. Spohn, in {\it Solids Far from Equilibrium Growth,
Morphology and Defects}, edited by C. Godreche (Cambridge University Press,
New York, 1990).

\item P. Meakin, {\it Fractal, Scaling and Growth Far from Equilibrium}
(Cambridge University Press, Cambridge, 1998).

\item M. Marsilli, A. Maritan, F. Toigoend, and J.R. Banavar, Rev. Mod. Phys.
{\bf 68}, 963 (1996).

\item M. Kardar, Physica A {\bf 281}, 295 (2000).

\item F. Family and T. Vicsek, J. Phys. A {\bf 18}, L75 (1985).

\item J. Krug, H. Kallabis, S.N. Majumdar, S.J. Cornell, A.J. Bray, and
C. Sire, Phys. Rev. E {\bf 56}, 2702 (1997).

\item For a review see, for example, S.N. Majumdar, Curr. Sci. {\bf 77},
370 (1999).

\item S.N. Majumdar and A.J. Bray, Phys. Rev. Lett. {\bf 86}, 3700 (2001).

\item S.O. Rice, Bell System Tech. J. {\bf 23}, 282 (1944); {\bf 24}, 46
(1945).

\item D.E. Newland, {\it An Introduction to Random Vibration, Spectral and
Wavelet Analysis} ( Longman, Harlow and Wiley, New York, 1993).

\item A. A. Masoudi, F. Shahbazi, J. Davoudi and M. R. Rahimi Tabar, Phys. Rev.
E {\bf 65}, 026132 (2002).

\item F. Shahbazi, S. Sobhanian, M.R. Rahimi Tabar, S. Khorram, G.R. Frootan,
and H. Zahed, J. Phys. A {\bf 36}, 2517 (2003).

\item A. Bahraminasab, S.M.A. Tabei, A.A. Masoudi, F. Shahbazi, and M.R.
Rahimi Tabar, J. Stat. Phys. {\bf 116}, 1521 (2004).

\item S.M.A. Tabei, A. Bahraminasab, A.A. Masoudi, S.S. Mousavi, and M.R.
Rahimi Tabar, Phys. Rev. E {\bf 70}, 031101 (2004).

\item R. Piessens, E. de Doncker, C. Uberhuber, D.K. Kahaner, {\it QUADPACK -
A Subroutine Package for Automatic Integration}, Springer Series in
Computational Mathematics (Springer, Berlin, 1983).

\item E.A. Novikov, Sov. Phys. JETP {\bf 20}, 1290 (1965).

\item J. Gar\'c\'ya-Ojalvo and J.M. Sancho, {\it Noise in Spatially Extended
Systems} (Springer-Verlag, New York, 1999).

\item  KR Sreenivasan, A. Prabhu and R. Narasimha, J. Fluid Mech.
{\bf 137}, 251 (1983).

\end{enumerate}
\newpage

 \begin{figure}[t]
 \begin{picture}(200,300)(0,0)
  \includegraphics{fig1.eps}
  \end{picture}
  \vspace{-3cm}
   \caption {Different time snapshots of the velocity fields of randomly
driven one-dimensional burgers equation
 }
\end{figure}
\begin{figure}
 \begin{picture}(200,300)(0,0)
  \includegraphics{fig2.eps}
  \end{picture}
   \vspace{-3cm}
 \caption{Plot of $\nu_{\alpha}^+$ vs $\alpha$ for the KPZ equation in the strong
coupling and before the emergence of sharp valleys for various
times, $t/t^*$ in $(2+1)-$dimensions(color online).}
\end{figure}

\begin{figure}
 \begin{picture}(200,300)(0,0)
  \includegraphics{fig3.eps}
  \end{picture}
   \vspace{-3cm}
\caption{Logarithmic plot of $N_{x_1}^+$ vs $t$ for the KPZ
equation in the strong coupling, and before the emergence of sharp
valleys in $(2+1)-$dimensions(color online).}
\end{figure}

\end{document}